\documentclass[aps,prd,reprint,twocolumn,showpacs,superscriptaddress,nofootinbib]{revtex4-1}  
\usepackage{tabularx}
\makeatletter
\def\hlinewd#1{%
\noalign{\ifnum0=`}\fi\hrule \@height #1 %
\futurelet\reserved@a\@xhline}
\makeatother
\usepackage{scrextend}
\usepackage{amsmath}
\usepackage{amssymb}
\usepackage{epsfig}
\usepackage{graphicx}
\usepackage{hyperref}
\usepackage[normalem]{ulem}
\usepackage{dcolumn}   % needed for some tables
\usepackage{slashed}
\usepackage{color}
\usepackage{rotating}
\usepackage[margin=0.9in,a4paper]{geometry}
\usepackage[table,xcdraw,dvipsnames]{xcolor}
\usepackage[utf8]{inputenc}
\usepackage{colortbl}
\usepackage[normalem]{ulem}
\usepackage{mathrsfs} 
\usepackage[version=4]{mhchem}
\usepackage{cancel}
\usepackage{xcolor}
\usepackage[utf8]{inputenc}
\usepackage{amsmath}
\usepackage{amssymb}
\usepackage{graphicx}
\usepackage{multirow}
\usepackage{hhline}
\usepackage[normalem]{ulem}
\usepackage{comment}
\usepackage{slashed}
\usepackage{pifont}

\usepackage{tikz-feynman}

\usepackage{array}
\newcolumntype{P}[1]{>{\centering\arraybackslash}p{#1}}

\definecolor{nicered}{rgb}{0.7,0.1,0.1}
\definecolor{nicegreen}{rgb}{0.1,0.5,0.1}
\definecolor{red}{rgb}{1.0, 0, 0}
\definecolor{CYAN}{rgb}{0, 1, 1} 

\newcommand{\magenta}[1]

\newcommand{\bdm}{\begin{displaymath}}
\newcommand{\edm}{\end{displaymath}}
\newcommand{\bea}{\begin{eqnarray}}
\newcommand{\eea}{\end{eqnarray}}
\newcommand{\ba} {\begin{equation}\begin{aligned}}
\newcommand{\ea} {\end{aligned}\end{equation}}

\newcommand{\pair}{a+{{}^{A}_{Z}X} \rightarrow {{}^{A}_{Z}X} + e^{+} + e^{-}}

\definecolor{nicered}{rgb}{0.7,0.1,0.1}
\definecolor{nicegreen}{rgb}{0.1,0.5,0.1}
\definecolor{red}{rgb}{1.0, 0, 0}
\definecolor{niceblue}{rgb}{0,0,0.8}
\definecolor{red}{rgb}{1.0, 0, 0}
\hypersetup{colorlinks,citecolor= nicegreen,linkcolor= nicered,urlcolor=nicered}

\def\gsim{\raise0.3ex\hbox{$\;>$\kern-0.75em\raise-1.1ex\hbox{$\sim\;$}}}
\def\lsim{\raise0.3ex\hbox{$\;<$\kern-0.75em\raise-1.1ex\hbox{$\sim\;$}}}
\def\mb[#1]{\mathbf{#1}}
\renewcommand{\bar}{\overline}

\definecolor{LightCyan}{rgb}{0.88,1,1}
\definecolor{piggypink}{rgb}{0.99, 0.87, 0.9}
\definecolor{applegreen}{rgb}{0.55, 0.71, 0.0}
\definecolor{darkpastelgreen}{rgb}{0.01, 0.75, 0.24}
\definecolor{green-yellow}{rgb}{0.68, 1.0, 0.18}
\newcommand{\beq}{\begin{equation}}
\newcommand{\eeq}{\end{equation}}
\newcommand{\beqa}{\begin{eqnarray}}
\newcommand{\eeqa}{\end{eqnarray}}
\newcommand{\eps}{\varepsilon}

\newcommand{\MeV}{{\, \rm MeV}}

\begin{document}

\title{Axion-induced pair production: a new strategy for axion detection
}

\author{Fernando Arias-Arag\'on} 
\email{fernando.ariasaragon@lnf.infn.it}
\affiliation{\small \it Istituto Nazionale di Fisica Nucleare, Laboratori Nazionali di Frascati, 00044 Frascati, Italy}
\author{Maurizio Giannotti} 
\email{mgiannotti@unizar.es}
\affiliation{\small \it Centro de Astropartículas y Física de Altas Energías, University of Zaragoza, Zaragoza, 50009, Aragón, Spain}
\affiliation{\small \it Department of Chemistry and Physics, Barry University, Miami Shores, 33161, FL, United States of America}
\author{Giovanni Grilli di Cortona} 
\email{giovanni.grilli@lngs.infn.it}
\affiliation{\small \it Istituto Nazionale di Fisica Nucleare, Laboratori Nazionali del Gran Sasso, Assergi, 67100, L’Aquila, Italy}
\author{Federico Mescia}
\email{federico.mescia@lnf.infn.it}
\affiliation{\small \it Istituto Nazionale di Fisica Nucleare, Laboratori Nazionali di Frascati, 00044 Frascati, Italy} 
\affiliation{\small \it On leave of absence from Universitat de Barcelona}

% ------------------------------------------------------
\begin{abstract}
\noindent
We revisit and update the axion-induced pair production process in a nuclear electric field mediated by the axion-electron coupling, $\pair$.  
This process emerges as one of the most efficient channels for detecting axions with energies above a few MeV in large underground detectors. 
It is particularly relevant for detecting axions produced in nuclear reactions, such as the $p+d~\rightarrow~{ }^3 \mathrm{He}~+~a(5.5\,\mathrm{MeV})$ reaction in the solar pp-chain, and for axions originating in supernovae.
Despite recent interest in detecting high-energy axions, the pair production process has received limited attention, even in scenarios where it is the dominant detection channel.
This study fills this gap by demonstrating that pair production is a highly effective detection mechanism for high-energy axions.
We apply our results to axions from supernovae and the solar 5.5 MeV line, recasting the current bounds of Borexino and comparing the detection capabilities of the JUNO and Hyper-Kamiokande detectors.

\end{abstract}

\maketitle

\tableofcontents

\section{Introduction}

Axions, hypothetical particles proposed to solve the strong CP problem in quantum chromodynamics (QCD)~\cite{Peccei:1977hh,Weinberg:1977ma,Wilczek:1977pj}, have drawn extraordinary interest in the field of particle physics and cosmology. 
Their feeble 
interactions with ordinary matter make them challenging to detect, yet their discovery could provide profound insights into the fundamental forces of nature and the composition of dark matter.
Traditional methods of axion detection have focused on processes such as axion-photon conversion in strong magnetic fields, which depends on the axion’s coupling to photons, as well as the axio-electric and axion inverse Compton processes, associated with the axion’s coupling to electrons. 
Each method presents its benefits and limitations, and the optimal detection method often depends on the axion flux targeted for detection. 

\begin{figure}[t]
\centering
\includegraphics[width=0.45 \textwidth,clip]{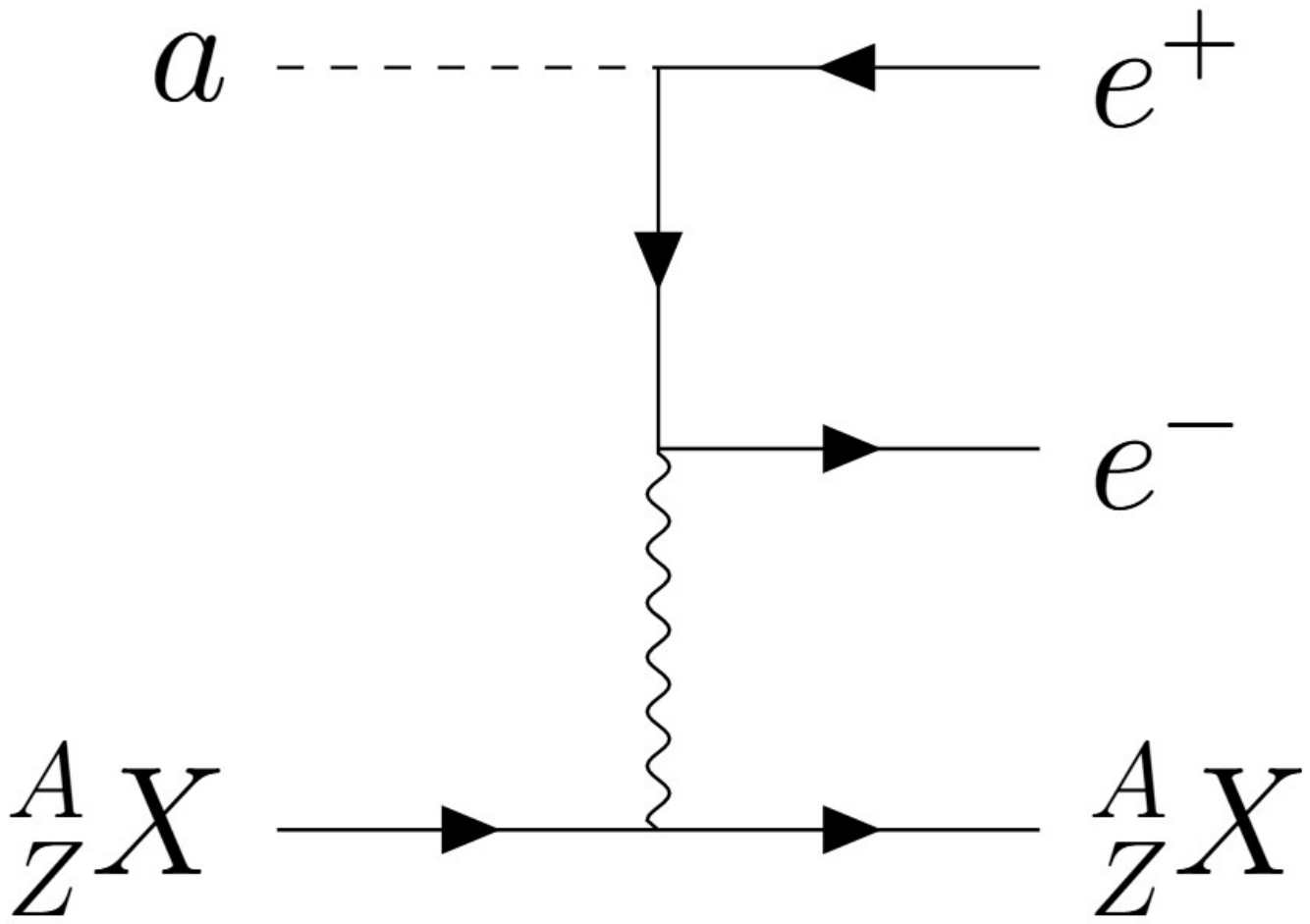}
\caption{External pair production induced by a pseudoscalar particle $a$ in the presence of a nucleus $_Z^AX$.
}
\label{fig:FeynPairProduction}       % Give a unique label
\end{figure}

In this paper, we study axion detection through the process  
\begin{align}
\label{eq:pair_process}
\pair \,,
\end{align}
induced by the axion-electron coupling, represented by the diagram in Fig.~\ref{fig:FeynPairProduction}. 
This is the axion equivalent of the famous Bethe-Heitler process~\cite{Bethe:1934za}.
While the process has been discussed in the literature, it has been largely overlooked in phenomenological and experimental applications, partly because of the lack of simple and practical expressions for the cross section on various targets — an issue we address in this work. 
As we shall show, this mechanism is particularly effective for detecting highly energetic axion fluxes, $E_a\gtrsim 10$ MeV, 
typically expected in core-collapse supernovae (SNe) or in nuclear reactions. 

The detection of MeV axions in large underground detectors has been considered in various publications. Examples include studies of the 5.5 MeV solar axion line from the proton-proton (pp) chain reaction
\begin{align}
\label{eq:pp-axions}
p + d \to {}^3\mathrm{He} + a(5.5\,\mathrm{MeV})
\end{align}
using Borexino data~\cite{Borexino:2012guz}, as well as sensitivity studies for the same axion line with future neutrino experiments~\cite{Lucente:2022esm}.
Both references considered the inverse Compton process, $a+e^{-} \rightarrow \gamma+e^{-}$, and the axio-electric effect (the axion equivalent of the photoelectric effect), as detection channels.\footnote{Their analyses also considered detection channels through the axion-photon coupling. Nevertheless, in this paper, we are interested only in the axion-electron coupling.}
However, these processes quickly become ineffective above a few MeV and lose their applicability entirely for detecting highly energetic axions, such as those expected from a supernova (SN) explosion, with energies that can exceed 100~MeV 
%
\begin{comment}
\begin{figure}[t]
\centering
\includegraphics[width=0.35 \textwidth,trim = 1 0.2 1 1,clip]{Figs/FeynmanPairProduction_v2.pdf}
\caption{External pair production induced by a pseudoscalar particle $a$ in the presence of a nucleus $_Z^AX$.\FM{to Fernando let's go back to your orginal diagrams}}
\label{fig:FeynPairProduction}       % Give a unique label
\end{figure}
\end{comment}
%

In this work, we present a detailed study of the axion pair production process in Fig.~\ref{fig:FeynPairProduction} (Eq.~\eqref{eq:pair_process}),  
and estimate the effect on the current bound of Borexino~\cite{Borexino:2012guz},
as well as the sensitivity at the future facilities Jiangmen Underground Neutrino Observatory (JUNO)~\cite{JUNO:2015zny} and Hyper-Kamiokande (Hyper-K)~\cite{Hyper-Kamiokande:2018ofw}. We limit the discussion to the mass range $m_a\ll2\,m_e$, a choice motivated by several reasons. First, this regime encompasses the standard QCD axion band. Second, in this regime, we can neglect axion decay en route to Earth and within the detector, avoiding complications that might divert attention from our primary results. Finally, in this limit, the matrix element calculations are simplified, enabling us to derive a phenomenological formula, Eq.~\eqref{eq:pair_prod_semian}, which relates the axion cross section to photon-induced external pair production in the electric field of nuclei.
As concrete examples, here we consider the following two axion sources: 
\begin{enumerate}
\item the 5.5 MeV axion flux from the solar pp-chain~\cite{Raffelt:1982dr,Borexino:2012guz}, Eq.~\eqref{eq:pp-axions}; 
\item the SN axion flux, produced through nucleon bremsstrahlung~\cite{Carenza:2019pxu} and pion Compton scattering~\cite{Lella:2022uwi,Lella:2023bfb}, in both trapping and free-streaming regime.
\end{enumerate}

In our study, we consider axions or axion-like particles (ALPs) coupled to electrons and nucleons through the following interaction terms: 
\begin{equation}
\label{eq:Lagrangian_int}
    \mathcal{L}_{\rm int} \supset -i g_{ae}\,a\,\bar{e}\gamma_5 e - i\,a\,\bar{N}\gamma_5 (g_{0aN}+\tau_3 g_{3aN})N,
\end{equation}
where $N=(n,p)^T$ refers to the  nucleon doublet and $\tau_3={\rm diag}(1,-1)$ is the third Pauli matrix. 
Using Eq.~\eqref{eq:Lagrangian_int}, the couplings to neutrons and protons can be expressed as $g_{an} = g_{0aN}+g_{3aN}$ and $g_{ap}~=~g_{0aN}-g_{3aN}$.
We ignore the axion-photon interaction as it is unnecessary for the examples we are considering here. 
Our main finding is that the pair production mechanism becomes competitive with the inverse Compton process at $E_a \approx 10$ MeV, as shown in  Sect.~\ref{sec:axion_production}.
Therefore, it remains somewhat subdominant, though not negligible, for the detection of the 5.5 MeV solar line, while it is by far the dominant detection channel in the case of SN axions.

The paper is organized as follows. 
We begin with a brief description of the two axion production mechanisms relevant to this work, outlined in Sec.~\ref{sec:axion_production}. 
Next, in Sec.~\ref{sec:axion_detection}, we discuss the axion detection mechanism associated with the axion-electron coupling and provide a detailed analysis of the axion pair production process. 
In Sec.~\ref{sec:experiments}, we review the experimental landscape, highlighting both the advantages and potential challenges. 
Finally, in Sec.~\ref{sec:results}, we summarize and discuss our results, with conclusions presented in Sec.~\ref{sec:conclusions}. 
A detailed calculation of the axion pair production cross section, including practical expressions for numerical analysis, is provided in Appendix~\ref{app:full}.

\section{Axion production}
\label{sec:axion_production}
Axions can be produced in a variety of ways, both in laboratory settings and in astrophysical environments.
Stars are an example of very effective axion sources. 
The hot stellar environment can produce axions through a variety of thermal processes (see, e.g., Ref.~\cite{DiLuzio:2020wdo, Antel:2023hkf, Caputo:2024oqc, Carenza:2024ehj} for reviews). 
In general, these processes produce axions with energies of the order of the stellar core, $\mathcal{O}$(1-10)~keV, for regular stars. 
However, regular stars, such as our own sun, can also produce more energetic axion fluxes through nuclear reactions and nuclear transitions. 
We discuss this below, in Sec.~\ref{sec:solar_axions}.
Another exception are SNe, see Sec.~\ref{sec:SNaxions}, with temperatures up to tens of MeV.

\subsection{Axions from nuclear reactions in the Sun}
\label{sec:solar_axions}

Axions can be produced in the Sun through non-thermal mechanisms involving nuclear reaction processes induced by the axion-nucleon couplings. 
Monochromatic fluxes arise from magnetic dipole transitions during the de-excitation of nuclei in the Sun~\cite{Donnelly:1978ty,Haxton:1991pu}, such as in the
$^{57}$Fe$^*\to ^{57}$Fe 
$+a(14.4$ keV) or 
$^{83}$Kr$^*\to ^{83}$Kr $+a(9.4$ keV), 
or from some nuclear reactions (see Tab. 4 in Ref.~\cite{Carenza:2024ehj} or Tab. 2 in Ref.~\cite{DiLuzio:2021qct} for more details and recent results).
In particular, the process in Eq.~\eqref{eq:pp-axions}, which represents the second step in the solar pp-chain but with an axion replacing the photon, turns out to be one of the most efficient production mechanisms induced by the axion-nucleon coupling~\cite{Massarczyk:2021dje}. 
The axion flux associated with this process is given by~\cite{Raffelt:1982dr,Borexino:2012guz} 
\begin{equation}
    \Phi_{a0}=(\Gamma_a/\Gamma_\gamma)\Phi_{\nu pp}\,,
\end{equation}
where $\Phi_{\nu p p}=6.0 \times 10^{10} \mathrm{~cm}^{-2} \mathrm{~s}^{-1}$ is the $p p$ solar neutrino flux and 
\begin{eqnarray}
\frac{\Gamma_a}{\Gamma_\gamma} &=& \frac{1}{2\pi\alpha} 
\left(\frac{|\vec k_a|}{|\vec k_\gamma|}\right)^3 \frac{1}{1+\delta^2} \left(\frac{g_{3aN}}{\mu_3-\eta}\right)^2
\nonumber\\
&\simeq& 0.54\, g_{3aN}^2 \left(\frac{|\vec k_a|}{|\vec k_\gamma|}\right)^3
\end{eqnarray}
denotes the probability of producing an axion rather than a photon.
Here, $\alpha$ denotes the fine-structure constant, while $|\vec k_a|$ and $|\vec k_\gamma|$ are the module of the axion and photon momenta, respectively (assumed equal in the limit of a massless axion). 
The others are nuclear parameters discussed in details, for example, in Ref.~\cite{Massarczyk:2021dje}.
For our specific reaction, the values $\delta = 0.82$ and $\eta=0$ apply, while the isovector magnetic moment is $\mu_3 = \mu_p - \mu_n = 4.77$.\footnote{We thank R. Massarczyk for clarifying a typo in Table II of Ref.~\cite{Massarczyk:2021dje}, which reported  $\delta=0$, $\eta=1$ instead of $\delta=0.82$, $\eta=0$. }
Substituting these parameters, we find the axion flux at earth as~\cite{Lucente:2022esm}
\begin{eqnarray}
     \Phi_a &=& \Phi_{a0}\,e^{-d_\odot/\ell_\mathrm{tot}}\\
     &\simeq& 3.23\times 10^{10}g_{3aN}^2 \biggl(\frac{|\vec k_a|}{|\vec k_\gamma|}\biggr)^3\,e^{-d_\odot/\ell_\mathrm{tot}}\,\,\mathrm{cm}^{-2}\mathrm{s}^{-1}   \nonumber,
\end{eqnarray}
where the exponential term $e^{-d_\odot/\ell_\mathrm{tot}}$, with 
$d_\odot~=~1.5\times 10^{13}$ cm the distance Sun-Earth and $\ell_\mathrm{tot}$ the total axion decay length, accounts for the possible decay of (sufficiently heavy) ALPs in their route from the sun. 
Though the axion could still couple with two photons at loop level—resulting in a finite (model-dependent) decay length which might suppress the flux—for simplicity, here we assume a vanishing axion-photon coupling and set the exponential prefactor to 1.

\subsection{SN axions}
\label{sec:SNaxions}

Core-collapse supernovae (SNe) are 
extraordinary environments for the production of feebly interacting particles. 
The violent gravitational collapse generates extreme temperatures and densities in the inner core~\cite{Janka:2006fh,Mirizzi:2015eza}, which in the cooling phase can reach the values $T\sim30-40\,\MeV$ and $\rho\sim10^{14}\,{\rm g\,cm^{-3}}$.
As we shall see, these conditions are ideal for the production of axions and axion-like particles. 
Indeed, a future galactic SN would offer a rare opportunity to probe the axion hypothesis as well as the physics of SNe and nuclear matter at extreme conditions.
Recent studies~\cite{Rozwadowska:2020nab} estimate the supernova rate to be $R = 1.63 \pm 0.46$ per century, 
with a distribution peaked around $10$ kiloparsecs (kpc) from Earth~\cite{Costantini:2005un} and concentrating in star-forming regions.

\begin{figure}[t]
\centering
\includegraphics[width=0.49 \textwidth,clip]{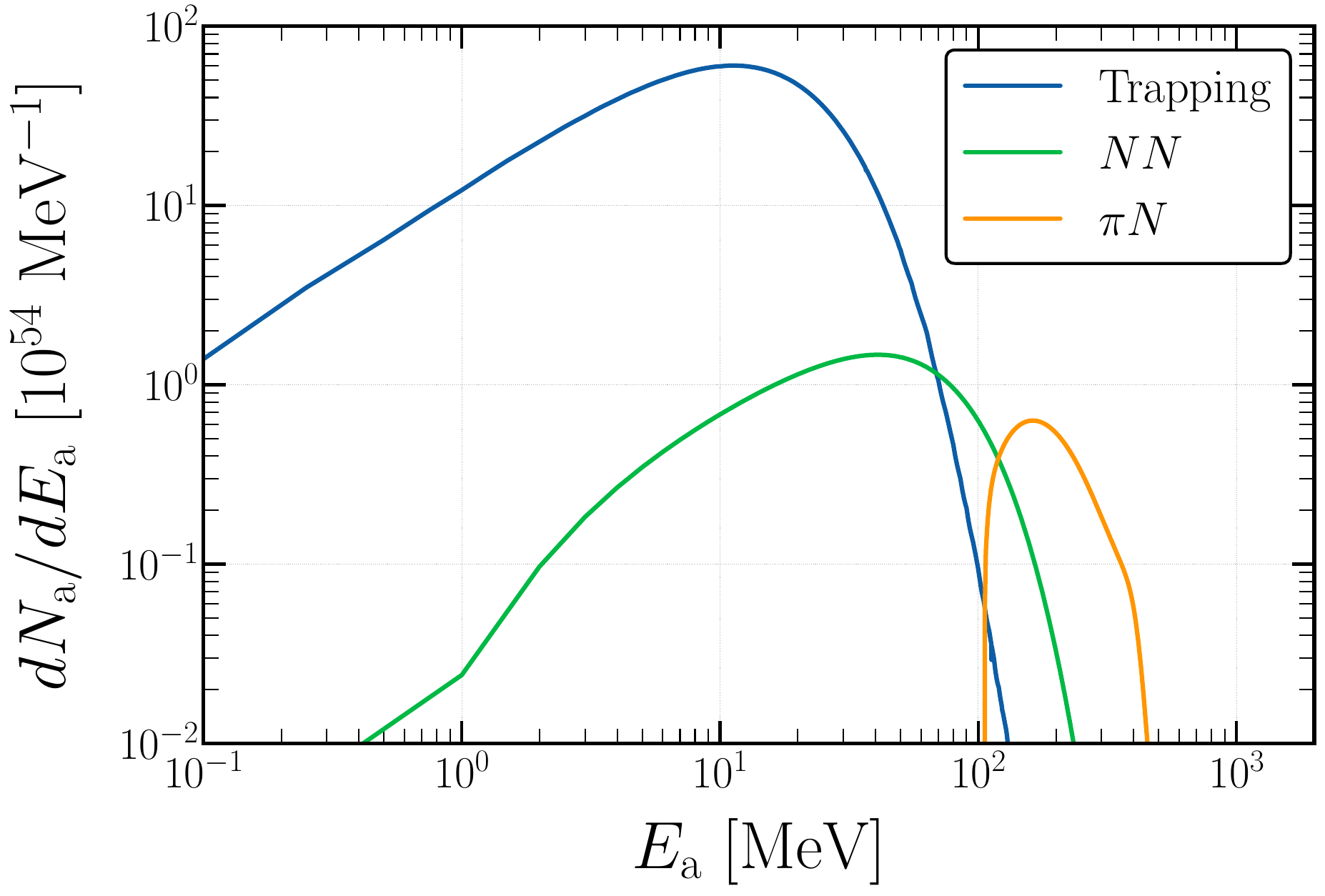}
\caption{Time-integrated axion production spectrum for the trapping regime (blue), $NN$ bremsstrahlung (green) and pion  conversion (yellow). The $NN$ and $\pi N$ contributions correspond to the free-streaming regime. 
The axion-nucleon couplings are fixed at $g_{ap}=5\times 10^{-10}$ and $g_{an}=0$. All curves scale as $g_{ap}^2$.}
\label{fig:SN_fluxes}       % Give a unique label
\end{figure}

If axions couple to nuclei, their production primarily occurs via nucleon-nucleon ($NN$) bremsstrahlung~\cite{Carenza:2019pxu,Lella:2022uwi,Lella:2023bfb,Lella:2024gqc}, $N+N \rightarrow N+N+a$ (see green curve of Fig.~\ref{fig:SN_fluxes}) 
and via pionic Compton processes, $\pi^-+p\to n+a$~\cite{Fore:2019wib,Carenza:2020cis,Ho:2022oaw}, shown in the yellow curve in Fig.~\ref{fig:SN_fluxes}.
The first of these processes is notoriously challenging to calculate, due to the complexity of describing the nucleon-nucleon interaction, and considerable effort has been made to improve the theoretical framework to provide a reliable description (see, in particular, Ref.~\cite{Carenza:2019pxu}).
The pion process, on the other hand, is easier to describe but depends critically on the unknown pion abundance in the SN core~\cite{Fore:2019wib,Carenza:2020cis}.
Specifically, by summing all contributions, the axion emission spectrum is expected to take the form~\cite{Lella:2022uwi,Lella:2023bfb} 
%%%%%%%%%%%%%%%%%%%%%%%%%%%%%%%%%%%%%%%%%%%% 
\begin{align} \label{eq
} \frac{dN_a}{dE}=  F_{NN}(E_a)+ F_{\pi N}(E_a)\,, 
\end{align} 
%%%%%%%%%%%%%%%%%%%%%%%%%%%%%%%%%%%%%%%%%%%% 
where the functions $F_{NN}$ and $F_{\pi N}$ represent the contributions of bremsstrahlung and pionic processes.\footnote{Notice that there are uncertainties in these processes related to the SN model, the description of the nucleon-nucleon interactions, medium effects in the SN, the pion abundance, etc. For more details, see e.g. Refs.~\cite{Carenza:2019pxu,Lella:2023bfb}.}
The pion scattering contribution, $F_{\pi N}$, adopted in our numerical analysis corresponds to a pion-to-nucleon fraction $Y_\pi =1\%$ in the inner core regions, a benchmark which agrees with recent theoretical estimates~\cite{Fore:2019wib}.
Interestingly, the fundamental question of the exact pion abundance in the SN core may only be resolved through a direct detection of SN axions, which motivates the ongoing experimental effort.

If the axion-nucleon couplings are sufficiently weak, ${g_{aN}\lesssim 10^{-8}}$, 
axions produced in the SN core would escape without being reabsorbed in the stellar matter. 
In this case, we talk about the \textit{free-streaming regime}.
A bound on these axions can be placed by constraining the effects of the additional axion cooling not to spoil the observed neutrino signal from SN 1987A. 
This results in ${g_{ap}\lesssim 8.5\times10^{-10}}$ in the case of axions interacting only with protons~\cite{Lella:2022uwi}. 
In the case of larger couplings, 
$g_{aN} \gtrsim 10^{-8}$, the SN environment becomes optically thick for the produced ALPs, preventing them from free-streaming out of the star~\cite{Lella:2023bfb}. 
This is dubbed the \textit{trapping regime}~\cite{Caputo:2022rca}, 
depicted by the blue curve of Fig.~\ref{fig:SN_fluxes}. 
Trapped axions are emitted with a typical thermal spectrum from a last scattering surface known as axionsphere, in analogy to the case of neutrinos. 
This flux can be constrained by the non-observation of associated events in the Kamiokande-II neutrino detector at the time of SN 1987A~\cite{Engel:1990zd,Lella:2023bfb}, resulting in the exclusion of 
$6 \times 10^{-10} \lesssim g_{a p} \lesssim 2.5 \times 10^{-6}$ for $m_a \lesssim 10 \,\mathrm{MeV}$~\cite{Lella:2023bfb}.

The interest in detecting supernova axions has increased in recent years and has led to various proposals, including searches with axion helioscopes, such as the International Axion Observatory (IAXO)~\cite{Ge:2020zww}, with the Large Hadron Collider (LHC)~\cite{Asai:2022pio}, or in neutrino water-Cherenkov detectors~\cite{Carenza:2023wsm,Carenza:2018jjc}. 
Additional proposals have also considered the opportunity to reveal SN ALPs 
with space born telescopes, specifically the Fermi LAT, in the case they convert in the galactic magnetic field~\cite{Meyer:2016wrm,Calore:2023srn,Lella:2024hfk}. 

Here, we show that the axion-induced pair production of electron and positron process offers also an opportunity to probe the SN axion spectrum, allowing to enlarge the reach in the parameter space with current and next-generation experiments. For our analysis, we calculated the SN axion spectrum in both free-streaming and trapping regime using the GARCHING group’s SN model SFHo-s18.8, provided in Ref.~\cite{SNarchive} and based on the {\tt PROMETHEUS-VERTEX} \cite{Rampp:2002bq} code, with the SFHo Equation of State (EoS) \cite{Hempel:2009mc,Steiner:2012rk}. 
The model consists in a $18.8M_\odot$ stellar progenitor \cite{Sukhbold:2017cnt}.

\section{Axion detection}
\label{sec:axion_detection}

Axions interactions with electrons lead to several detection channels. 
The ones we are interested in, in this work, are
\begin{enumerate}
    \item inverse Compton scattering, $a+e^- \to \gamma + e^-$;
    \item the axio-electric effect $a+e^-+Ze \to e^- + Ze$; 
    \item axion-induced pair production in the electric field of nuclei, $\pair$. 
\end{enumerate}
Below, we provide a brief overview of each and discuss axion pair production in more detail, as it is considerably less explored in the literature.
% 
\begin{comment}
\begin{figure*}[ht!]
\centering
\includegraphics[width=0.25 \textwidth,trim = 10 90 10 90,clip]{Figs/FeynmanInverseCompton.pdf}
\includegraphics[width=0.35 \textwidth,trim = 10 170 70 180,clip]{Figs/FeynmanAxioelectric.pdf}
\includegraphics[width=0.25 \textwidth,trim = 1 0.2 1 1,clip]{Figs/FeynmanPairProduction_v2.pdf}
\caption{External pair production induced by a pseudoscalar particle $a$ in the presence of a nucleus $_Z^AX$.}
\label{fig:detection_feyn}       % Give a unique label
\end{figure*}
\end{comment}

\subsection{Inverse Compton scattering}

The total cross section for the process where an axion scatter on the atomic electrons of the target, $a+e^-\to e^- + \gamma$, is~\cite{Zhitnitsky:1979cn,Avignone:1988bv,Donnelly:1978ty}
\begin{eqnarray}
\label{eq:inverse_compton}
    \sigma_\textit{IC}&=&  \frac{g_{ae}^2\alpha}{8 m_e^2 |\vec k_a|} \Bigg[\frac{2m_e^2(m_e + E_a)y}{(m_e^2 + y)^2} \nonumber\\
    &+& \frac{4 m_e(m_a^4 + 2 m_a^2 m_e^2 - 4 m_e^2 E_a^2)}{y(m_e^2 + y)} \\
&+& \frac{4 m_e^2 |\vec k_a|^2 + m_a^4}{|\vec k_a| y} \ln{\frac{m_e + E_a + |\vec k_a|}{m_e + E_a - |\vec k_a|}} \Bigg] \nonumber
\end{eqnarray}
where $y=2m_e E_a + m_a^2$, $k_a$ is the momentum of the axion, $E_a$ its energy and $m_e$ the electron mass. 
This process produces an observable signature in the electron recoil and outgoing photon final states. 
For $m_a\lesssim 2$ MeV, the phase-space contribution to the cross section in equation~\eqref{eq:inverse_compton} is approximately independent of the axion mass and the cross section becomes $\sigma_{IC}\simeq 4.3\times 10^{-25}\,g_{ae}^2$ cm$^2$.

\subsection{Axio-electric effect}
In analogy with the photo-electric effect, in the axio-electric effect the axion is absorbed by the atom and an electron with an energy equal to the difference between the axion energy and the electron binding energy is emitted~\cite{Dimopoulos:1985tm,Smith:1988kw,Rosenberg:2000wb,Pospelov:2008jk,Arisaka:2012pb}. The cross section is
\begin{equation}
    \sigma_{ae} = \sigma_{\rm pe}\frac{g_{ae}^2}{\beta_a}\frac{3E_a^2}{16\pi\alpha m_e^2}\biggl(1-\frac{\beta^{2/3}_a}{3}\biggr)\,,
\end{equation}
where $\sigma_\mathrm{pe}$ is the photoelectric cross section of the target, that can be obtained from the \textit{Photon Cross Sections Database} in~\cite{xcom}, and $\beta_a~=~|\vec k_a|/E_a$. The photo-electric cross section $\sigma_\mathrm{pe}$ is proportional to $Z^5$, making the axio-electric effect the main (solar or galactic) axion detection channel in direct detection experiments with high-Z targets~\cite{LUX:2017glr,PandaX:2017ock,XENON:2021qze,DarkSide:2022knj}. However, the steep decrease of the cross section at higher energy makes this channel ineffective for axions with $E_a\gtrsim50$ keV.

\begin{comment}
\subsection{Axion decay}
\mg{I dont think we ever need this, unless we want to account for the possible decay of heavy ALPs on route to Earth.}
Axions with mass $m_a>2 m_e$ may decay into $e^+e^-$ pairs with a decay length 
\begin{eqnarray}
    \ell_e=\frac{\gamma v}{\Gamma_{a\to e^+e^-}}\,,
\end{eqnarray}
where $\Gamma_{e^+e^-}= m_a\,{g_{ae}}/{8\pi}  \sqrt{1-4\,m_e^2/m_a^2}$.
\end{comment}

\subsection{Axion-induced external pair production}

Finally, an axion can produce electron-positron pairs in the electric field of nuclei. The relevant cross section was calculated in Refs.~\cite{Kim:1982xb,Kim:1984ss,Blumlein:1991xh,Zhitnitsky:1979cn,Bardeen:1978nq}. 
This process is analogous to the photon-lepton pair production in the Standard Model, known as the Bethe-Heitler process, and can be evaluated using the form factor formalism from Ref.~\cite{Tsai:1973py}.
The differential cross section is given by 
\begin{equation}
  \!\!\!  \dfrac{d\sigma_{aee}}{d\cos\theta_+d\cos\theta_-d E_-d\phi} = - \frac{|\mathcal{M}|^2  \, |\vec{\ell}_+| \, |\vec{\ell}_-| }{512\, \pi^4\, m_X^2\, |\vec k_a| },
\end{equation}
where $E_a$, $E_+$, $E_-$ are the axion, positron and electron energy, respectively. 
The axion momentum is denoted by $k_a$, while $m_X$ is the mass of the nucleus. Energy conservation imposes the relation $E_a=E_++E_-$ between the initial state axion, and final state positron and electron energy, as long as nuclear recoil is neglected. The integration range is $\theta_\pm \in [0,\,\pi]$, $\phi\in [0,\,2\pi]$ and $E_-\in [m_e,\,E_a - m_e]$. The modulus squared of the matrix element can be written as
\begin{equation}
    |\mathcal{M}|^2 = \frac{e^4 g_{ae}^2}{t^2} L_{\mu\nu}H^{\mu\nu}\,F_A^2(t). 
    \label{eq:Mq}
\end{equation}
The $1/t^{2}$ factor comes from the propagator of the photon and it is
\begin{eqnarray}
  \!\!\!  t&=&(k_a-\ell_+-\ell_-)^2\\ 
    &=& m_a^2 + 2 m_e^2 -2\left( k_a\cdot \ell_+ + k_a\cdot \ell_- - \ell_-\cdot \ell_+\right)\,,\nonumber
\end{eqnarray}
where $k_a$ and $\ell_{\pm}$ are the four-momenta of the axion and positron/electron, respectively.
The leptonic and hadronic tensors are defined as 
\begin{eqnarray}
   \!\!\!\!L_{\mu\nu} &=& \mathrm{Tr}\biggl[ (\slashed{\ell}_-+m_e) 
    \biggl( \gamma_\mu 
    \frac{\slashed{k}_a-\slashed{\ell}_++m_e}{m_a^2-2(k_a\cdot\ell_+)} \gamma_5 \nonumber\\
    && \quad+\gamma_5 \frac{\slashed{\ell}_--\slashed{k}_a+m_e}{m_a^2-2(k_a\cdot\ell_-)} \gamma_\mu \biggr)
    (\slashed{\ell}_+-m_e)\nonumber\\
    &&\quad\times\biggl(\gamma_5 \frac{\slashed{k}_a-\slashed{\ell}_+ + m_e}{m_a^2-2(k_a\cdot\ell_+)} \gamma_\nu \\
    &&\quad+ \gamma_\nu \frac{\slashed{\ell}_- - \slashed{k}_a + m_e}{m_a^2-2(k_a\cdot\ell_-)}\gamma_5  \biggr) \biggr],\nonumber\\
    \!\!H^{\mu\nu} &=& \mathrm{Tr}\biggl[ \left( \slashed{p}_2 + m_X \right) \gamma^\mu \left( \slashed{p}_1 + m_X \right) \gamma^\nu \biggr].\nonumber
\end{eqnarray}
The atomic form factor is~\cite{Tsai:1973py,Kim:1973he}
\begin{equation}
    F_A^2(t) = Z^2\left[ \frac{a^2_Z t}{1+a^2_Z t}\frac{1}{1+t/d_A} \right]^2,
    \label{eq:FF}
\end{equation}
which includes the electron screening of the nuclear charge. In equation~\eqref{eq:FF}, $a_Z=111\,Z^{-1/3}/m_e$, $d_A=0.164\,A^{-2/3}$ GeV$^2$, and $A$ is the atomic mass number. More details on the computation can be found in Appendix~\ref{app:full}. 

The axion-induced external pair production in the nuclear electric field can be related to the analogous process of external pair production by a photon in the nuclear electric field. Comparing the two computations in the limit of $m_a\ll m_e$, we find
\begin{equation}
    \sigma_{aee}(E_a)\simeq 2.6\,\frac{g_{ae}^2}{4\pi\,\alpha} \sigma_{\gamma ee}(E_a),
    \label{eq:pair_prod_semian}
\end{equation}
where $\sigma_{\gamma ee}$  is the cross section for photon-induced external pair production in the electric field of nuclei, which can be obtained from the database~\cite{xcom}.

\begin{figure}[t]
\centering
\includegraphics[width=0.49 \textwidth,clip]{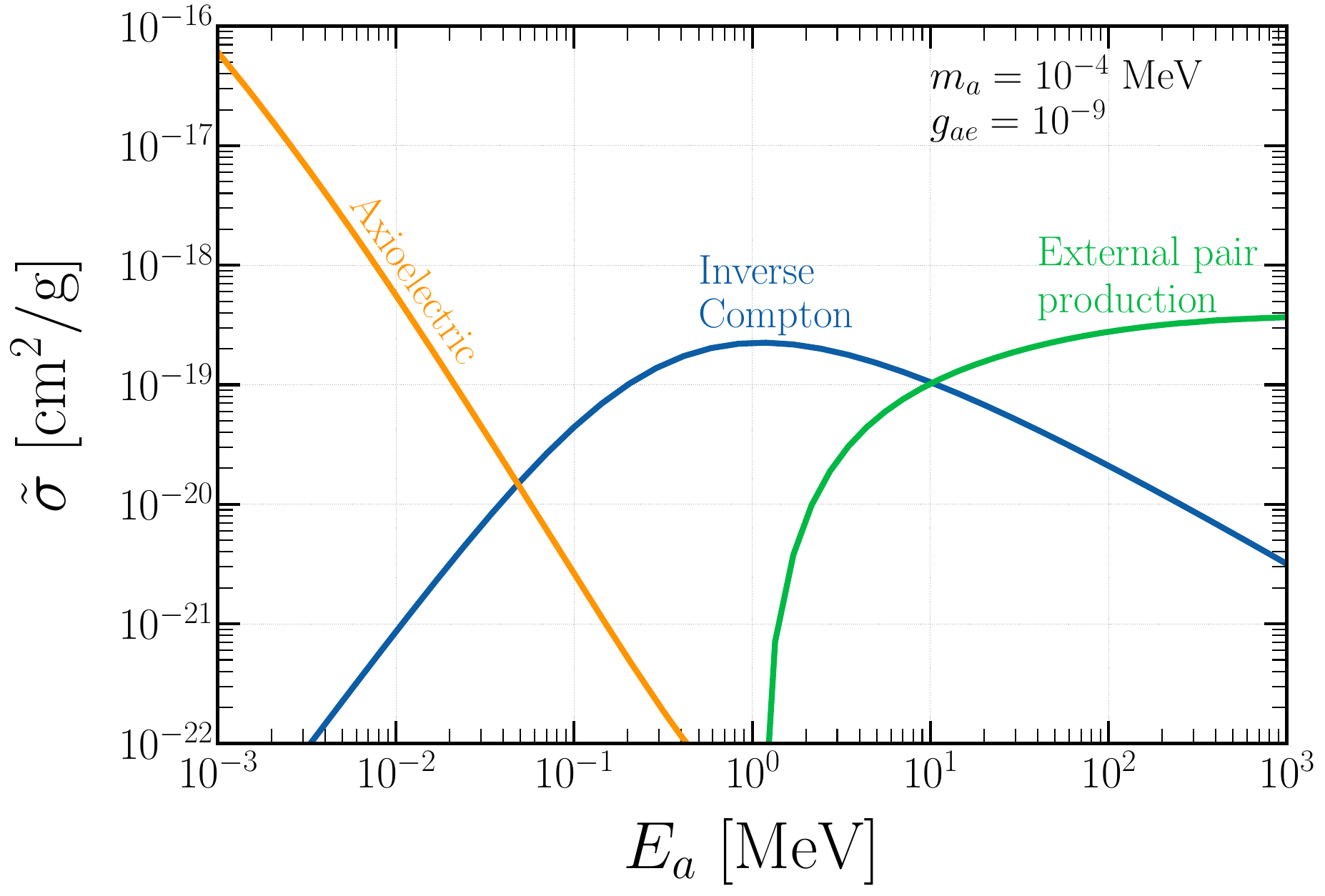}
\caption{Specific cross section $\tilde{\sigma}$ in cm$^2$/g for the axioelectric (yellow), the inverse Compton (blue) and the external pair production (green) processes in LAB (C$_{19}$H$_{32}$) for $g_{ae}=10^{-9}$ and $m_a=10^{-4}$ MeV. \label{fig:xscomparison}}       % Give a unique label
\end{figure}

Figure~\ref{fig:xscomparison} shows the comparison between the cross sections for the axio-electric effect (orange), the  inverse Compton scattering (blue) and the external pair production channel (green) for an axion with a coupling to electrons of $g_{ae}=10^{-9}$ and a mass of $m_a=10^{-4}$ MeV, interacting in LAB ($C_{19}H_{32}$). 
The three different processes dominate in different ranges of axion energies. 
For $E_a \lesssim 5\times 10^{-2}$~MeV the axio-electric effect dominates; the inverse Compton scattering is the leading process between $5\times 10^{-2} \lesssim E_a / \mathrm{MeV} \lesssim 10$, while for $E_a \gtrsim 10$ MeV signal from external pair production will be dominant. 
This figure allows us to foresee that the pair production mechanism will be subdominant in the case of solar axions, although it will give a non-negligible contribution.
On the other hand, for axions produced by SN it will dominate the detection.

\section{Experiments}
\label{sec:experiments}
In this section, we will give a brief description of the detectors that we are interested in, detailing their composition, active mass and their energy resolution. More details can be found in Ref.~\cite{Borexino:2012guz,JUNO:2015zny,Hyper-Kamiokande:2018ofw}.

\subsection{Borexino}

Borexino was a spherical scintillator detector with a 278-ton active mass of pseudocumene ($C_9H_{12}$), doped with 1.5 g/L of PPO ($C_{15}H_{11}NO$), now in the decommissioning phase. This active mass was housed in a nylon vessel, surrounded by two pseudocumene buffers separated by a nylon membrane. The entire setup was contained within a stainless steel sphere and a water tank, which shields against external radiation and acts as a Cherenkov muon detector. Borexino's energy resolution is given by $\sigma/E = (0.058 + 1.1 \times 10^{-3} E)/\sqrt{E}$, with $E$ in MeV~\cite{Borexino:2012guz}.

\subsection{JUNO}

JUNO's detector consists in a central water-Cherenkov detector and a muon tracker. The water-Cherenkov detector is filled with a liquid scintillator fiducial volume of up to $20$ kton of linear alkylbenzene (LAB, $C_{19}H_{32}$), doped with $3$ g/L of 2.5-diphenylosazole (PPO) and $15$ mg/L of p-bis-(o-methylstyryl)-benzene (bis-MSB) and with an excellent energy resolution of $\sigma/E = 0.03/\sqrt{E}$,  with $\sigma$ and $E$ in MeV~\cite{JUNO:2015zny}. The liquid scintillator is contained in a spherical vessel of radius 17.7 m, surrounded by  more than $5\times 10^4$ photomultipliers and it has a density of $0.859$ g/ml. The fiducial volume considered depends on the energy, in order to reduce the background and detect $^8$B solar neutrinos: for energies $2 < E/\mathrm{MeV} \leq 3$ the fiducial mass is $7.9$ kton, for $3 < E/\mathrm{MeV} \leq 5$ the fiducial mass is of $12.2$ kton, and for $E>5$ MeV the fiducial mass is of $16.2$ kton. More details about analysis cuts to reduce the background can be found in Ref.~\cite{JUNO:2020hqc}. JUNO is expected to start taking data in the second half of 2025~\cite{reuters2024undergroundlab}.

\subsection{Hyper-Kamiokande}

Hyper-Kamiokande is a next generation neutrino observatory planned to be installed in Japan, near Kamioka, and expected to start in 2027~\cite{Hyper-Kamiokande:2018ofw}. The detector is a large, cylindrical water-Cherenkov detector with 187 kton fiducial mass~\cite{Hyper-Kamiokande:2021frf}. We use the same energy resolution of the latest Super-Kamiokande analysis for solar neutrinos, $\sigma/\mathrm{MeV}~=~-0.05525 + 0.3162 \sqrt{E} + 0.04572 E$~\cite{Super-Kamiokande:2023jbt}. 

\subsection{Detection rates}
Let us now introduce some useful details for the discussion of experimental bounds and sensitivities. {In particular, we will define our signal events and describe the statistical analyses performed in order to set limits by recasting existing bounds and giving projections for future searches.}

For solar axions, we evaluate their constraints at Borexino and explore their detection potential at future experiments such as JUNO and Hyper-K. 
Additionally, we consider the sensitivity of these latter two experiments to detect axions from Supernova sources.

An axion interaction with the detector results in an expected number of signal events given by:
\begin{equation}
    \label{eq:Sexp}
    S_\mathrm{expected} = \epsilon\, N_T\,T\, \Phi_a \otimes \sigma\, ,
\end{equation}
where $\epsilon$, $T$ and $N_T$  depend on the experimental setup of the detector.
Specifically, $\epsilon$ is the detector efficiency, $T$ is the measurement time (which, for the SN case, is taken to be 10 s corresponding to the interval of the burst), and $N_T$ represents the number of target particles in the detector (electrons for inverse Compton scattering and the axio-electric effect, or nuclei for external pair production). The short notation $\Phi_a \otimes \sigma$ represents the initial axion flux integrated with the signal cross section $\sigma$ for a given process. 

Following the likelihood analysis of Ref.~\cite{Lucente:2022esm}, we compute the $90\%$ limit for solar axions at Borexino, and the sensitivity of JUNO, and Hyper-K. 
For Borexino, we use the limit of  $S_\mathrm{lim} = 6.9$ events over 536 days with a fiducial mass of 278 tons of pseudocumene ($C_9H_{12}$), taken from the experimental analysis of Ref.~\cite{Borexino:2012guz}.

For JUNO {and Hyper-K}, we estimated the number of background events from Ref.~\cite{JUNO:2020hqc}. In particular, we use the $10$ years spectra from $B^8$, $hep$, reactor neutrinos, denoted as $N_i^\mathrm{sb}$, and from radioactive backgrounds, $N_i^\mathrm{rb}$. We obtain the detector sensitivity employing a $\chi^2$ function
\begin{eqnarray}
    \chi^2&=&2\,\sum_i\left( N_i^\mathrm{sig} - N_i^\mathrm{bkg} + N_i^\mathrm{bkg}\,\log\frac{N_i^\mathrm{bkg}}{N^\mathrm{sig}_i}\right) \nonumber\\
    &+&\left(\frac{\epsilon_\mathrm{sb}}{\sigma_\mathrm{sb}}\right)^2+\left(\frac{\epsilon_\mathrm{rb}}{\sigma_\mathrm{rb}}\right)^2,
    \label{eq:chi2}
\end{eqnarray}
where $N_i^\mathrm{bkg}=N_i^\mathrm{sn} + N_i^\mathrm{rb}$ are the number of events in the background for the $i^\mathrm{th}$ energy bin, $N_i^\mathrm{sig} = (1+\eps_\mathrm{sb})\,N_i^\mathrm{sb} + (1+\eps_\mathrm{rb})\,N_i^\mathrm{rb} +  N_i^\mathrm{axions}$. 
In Eq.~\eqref{eq:chi2}, $\eps_\mathrm{sb}$ and $\eps_\mathrm{rb}$ are the nuisance parameters corresponding to solar and radioactive background normalization, with uncertainties $\sigma_\mathrm{sb}=0.05$ and $\sigma_\mathrm{rb}=0.15$ respectively. 
The axion signal contribution is $N_i^\mathrm{axions}=S\,\mathcal{G}(E_i,\bar{E}_a,\bar{\sigma}_a)$, where $S$ parametrizes the expected axion intensity, and $\mathcal{G}$ is a Gaussian function centered at $\bar{E}_a=5.49$ MeV and with a width of $\bar{\sigma}_a$, given by the JUNO or Hyper-K detector energy resolution at $5.49$ MeV. 
For JUNO, we find $S_\mathrm{lim} = 104$ events at $90\%$ CL over 10 years with JUNO's fiducial mass of linear alkylbenzene (LAB) ($C_{19}H_{32}$)~\cite{JUNO:2015zny}. Hyper-K's limit is  $S_\mathrm{lim} = 3.5\times10^3$ events in a 187 kton water detector~\cite{Hyper-Kamiokande:2018ofw,Hyper-Kamiokande:2021frf}.\footnote{For JUNO, the difference with respect to Ref.~\cite{Lucente:2022esm} is due to the different background used, while for Hyper-K because of the different fiducial mass and energy resolution.}

For the SN axion analyses, on the other hand, we use a different strategy. The dominant background in the $10$ s burst is given by the SN neutrino produced. The background spectra, for both JUNO and Hyper-K, will be dominantly at energies below $100$ MeV. We then require the detection of $S_\mathrm{lim} = 3$ axion events above the cut-off energy $E_\mathrm{cut-off}=100$ MeV, where the background is negligible.\footnote{This is true for both the $200$ pc and $10$ kpc benchmark distances of our analysis, and it may be modified or refined. Since we are only interested on the relative strength of the inverse Compton or pair production detection channels, a more refined analysis is beyond the scope of this study.} 

In the following section, we use these limits to set bounds on the couplings requiring from different experiments that $S_\mathrm{expected} \leq S_\mathrm{lim}$. 

\section{Results}
\label{sec:results}

In this section, we present the results obtained from our analysis for solar and SN axions {in all the different experiments already described. We will give, for each case, the expected number of events as a function of the axion couplings. This result is independent of the axion mass for masses roughly below the MeV. Then, using the limit number of events $S_\mathrm{lim}$ obtained in the previous section for each experiment, we will show the reach in the axion couplings.} 

\subsection{Solar axions}

The expected number of events in Borexino due to inverse Compton scattering is calculated based on the number of electrons $N_e=9.17\times10^{31}$ in the target, the exposure time of $536$ days, and the experimental detection efficiency $\epsilon=0.358$,\footnote{For pair-production we used the same efficiency, as an estimate. A complete analysis of this efficiency is outside the scope of this work.} as provided by the Borexino Collaboration~\cite{Borexino:2012guz}. The inverse Compton cross section for $m_a\lesssim 1$ MeV is $\sigma_\mathrm{IC} \simeq 4.3\times 10^{-25}\, g_{ae}^2$ cm$^2$, resulting in 
\begin{equation}
    S_\mathrm{expected} \simeq 2.1\times 10^{25}\, g_{3aN}^2\,g_{ae}^2,
\end{equation}
and a bound of $|g_{3aN}\,g_{ae}|\lesssim 5.7\times 10^{-13}$.

Similarly, the number of events in Borexino due to external pair production by an axion in the electric field of the nuclei for $m_a \leq 1$ MeV is calculated. The number of carbon and hydrogen atoms in Borexino's inner vessel is $N_C=1.25\times 10^{31}$ and $N_H=1.67\times 10^{31}$, respectively. The pair production cross section at $E_a \sim 5.5$ MeV is $\sigma_C=1.27\times 10^{-24}\,g_{ae}^2$ cm$^2$ for carbon and $\sigma_H=3.54\times 10^{-26}\,g_{ae}^2$ cm$^2$ for hydrogen. This yields an expected number of events of 
\begin{equation}
    S_\mathrm{expected} \simeq 8.8\times 10^{24} \, g_{3aN}^2\,g_{ae}^2\,,
\end{equation}
providing a bound of $|g_{3aN}\,g_{ae}|\lesssim 8.8\times 10^{-13}$ solely from pair production. 

Consequently, the total number of expected events in the Borexino detector is \begin{equation}
    S_\mathrm{expected}\simeq 3.0\times 10^{25}\,g_{3aN}^2\,g_{ae}^2,
\end{equation}
which leads to a slightly improved bound of $|g_{3aN}\,g_{ae}|\lesssim4.8\times 10^{-13}$, compared to results from the inverse Compton scattering alone~\cite{Borexino:2012guz}.

In JUNO's fiducial volume, the number of electron targets is $N_e\sim 5.5\times 10^{33}$, and we assume $\epsilon=1$ since only the fiducial volume is used. Over $10$ years, the number of expected events from inverse Compton conversion in JUNO would be \begin{equation}
S_\mathrm{expected}\simeq 2.4\times 10^{28}\,g_{3aN}^2\,g_{ae}^2,
\end{equation}
allowing sensitivity down to $|g_{3aN}\,g_{ae}|\sim 6.6\times 10^{-14}$. 

For axion-induced external pair production in JUNO, the  number of events depends on the number of carbon $N_C=6.13\times10^{32}$ and hydrogen $N_H=1.02\times 10^{33}$ targets, resulting in an expected number of events 
\begin{equation}
S_\mathrm{expected}\simeq 9.6\times 10^{27}\,g_{3aN}^2\,g_{ae}^2.
\end{equation}
The corresponding sensitivity for pair production alone reaches $|g_{3aN}\,g_{ae}| \sim 10^{-13}$. 

Combining both processes, the expected number of events increases to 
\begin{equation}
    S_\mathrm{expected}\simeq 3.4 \times 10^{28}\,g_{3aN}^2\,g_{ae}^2,
\end{equation}
and the sensitivity further improves to $|g_{3aN}\,g_{ae}| \sim 5.6 \times 10^{-14}$. 

For Hyper-K, the corresponding bounds are worsened with respect to JUNO by about a factor 2. 
We refer to Tab.~\ref{tab:Summary} for the corresponding sensitivities in this case.
\begin{figure}[t]
\centering
\includegraphics[width=0.49 \textwidth,clip]{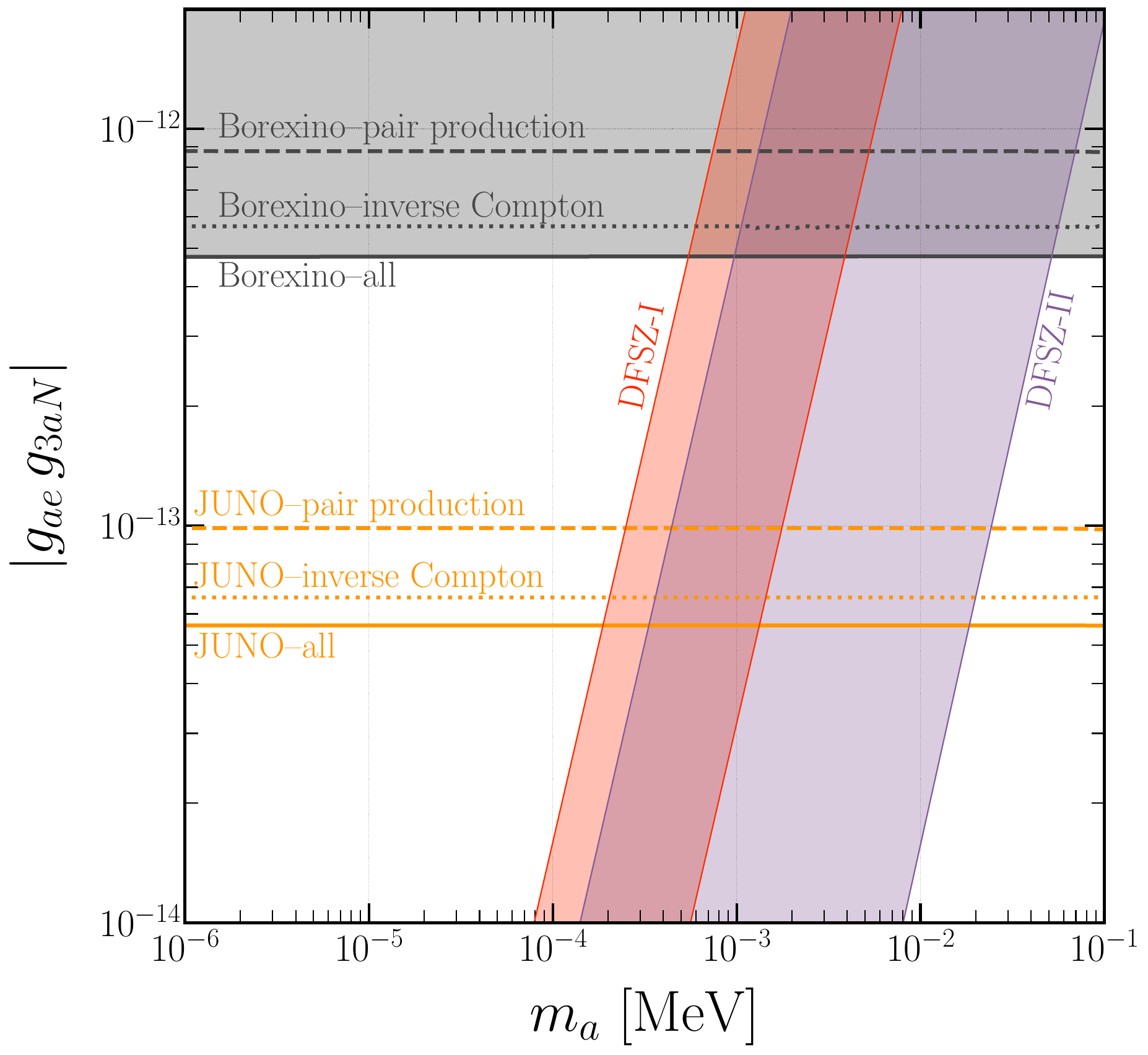}
\caption{Exclusion region in the $(|g_{3aN}\, g_{ae}|,\,m_a)$ plane at $90\%$ C.L. The gray lines represent the solar axion bound from Borexino from external pair production (dashed), inverse Compton (dotted) and their combination (solid). The yellow curves represent the sensitivity of JUNO. The red and purple bands show the preferred regions for the DFSZ-I and DFSZ-II models.\label{fig:solar_final}}       % Give a unique label
\end{figure}
Fig.~\ref{fig:solar_final} presents the results for Borexino and JUNO, displaying the couplings $|g_{ae}\,g_{3aN}|$ as a function of the axion mass $m_a$. The red and purple bands represent the preferred regions for the DFSZ-I and DFSZ-II models, respectively~\cite{Zhitnitsky:1980tq,Dine:1981rt,DiLuzio:2020wdo,GrillidiCortona:2015jxo}.
The gray curves denote the bounds from Borexino: the dashed curve shows the bound from external pair production by an axion alone, the dotted curve from inverse Compton scattering, and the solid curve from both processes. 
Similarly, the yellow dashed, dotted and solid curves denote the sensitivity reach of a $10$-year measurement at JUNO for the pair production, inverse Compton scattering, and their combination, respectively.
As shown in Fig.~\ref{fig:solar_final}, adding the external pair production process marginally improves both the bound from Borexino and the sensitivity of JUNO (and Hyper-K, see Tab.~\ref{tab:Summary}). Different results are expected for SN fluxes, as they produce more energetic axions.

\begin{comment}
\ggdc{The part below should be modified taking into account the new fiducial mass of $187$ kton and the new Slim of 3500 events.}
    the inverse Compton scattering event count is based on the number of electron targets in the $374$ kton of water target, $N_e\simeq 1.25\times 10^{35}$, yielding an expected event count over 10 years of 
\begin{equation}
S_\mathrm{expected} \simeq 5.48 \times 10^{29}\,g_{3aN}^2\,g_{ae}^2,
\end{equation}
and a sensitivity down to $|g_{3aN}\,g_{ae}| \sim 1.34 \times 10^{-13}$. 

On the other hand, the number of oxygen and hydrogen targets in the detector are $N_O=1.25\times 10^{34}$ and $N_H=2.5\times 10^{34}$, respectively. The pair production cross sections are $\sigma_O=2.27\times 10^{-24}\,g_{ae}^2$ cm$^2$ and $\sigma_H=3.54\times 10^{-26}\,g_{ae}^2$ cm$^2$, giving an expected number of events \begin{equation}
S_\mathrm{expected} \simeq 2.98\times 10^{29} \,g_{3aN}^2\,g_{ae}^2,
\end{equation}
and sensitivity down to $|g_{3aN}\,g_{ae}| \sim1.82\times 10^{-13}$. 

Combining both processes in Hyper-K enhances the expected number of events to
\begin{equation}
    S_\mathrm{expected} \simeq 8.5\times 10^{29} \,g_{3aN}^2\,g_{ae}^2,
\end{equation}
and improve the sensitivity to $|g_{3aN}\,g_{ae}| \sim 1.08 \times 10^{-13}$, which is about twice as weak as JUNO's potential sensitivity.
\end{comment}

\subsection{SN axions}

\begin{figure*}[t]
\centering
\includegraphics[width=0.49 \textwidth,clip]{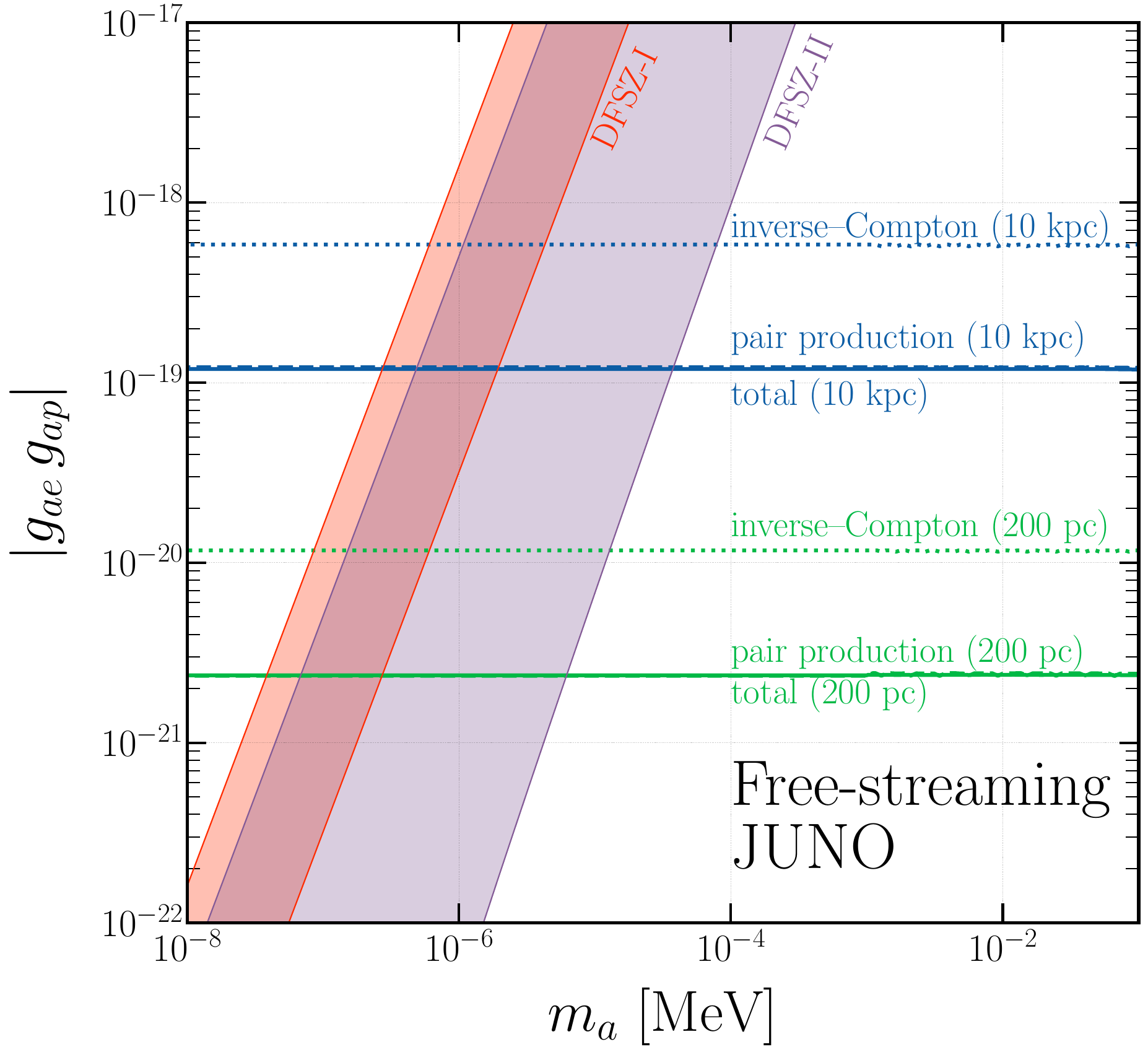}
\includegraphics[width=0.49 \textwidth,clip]{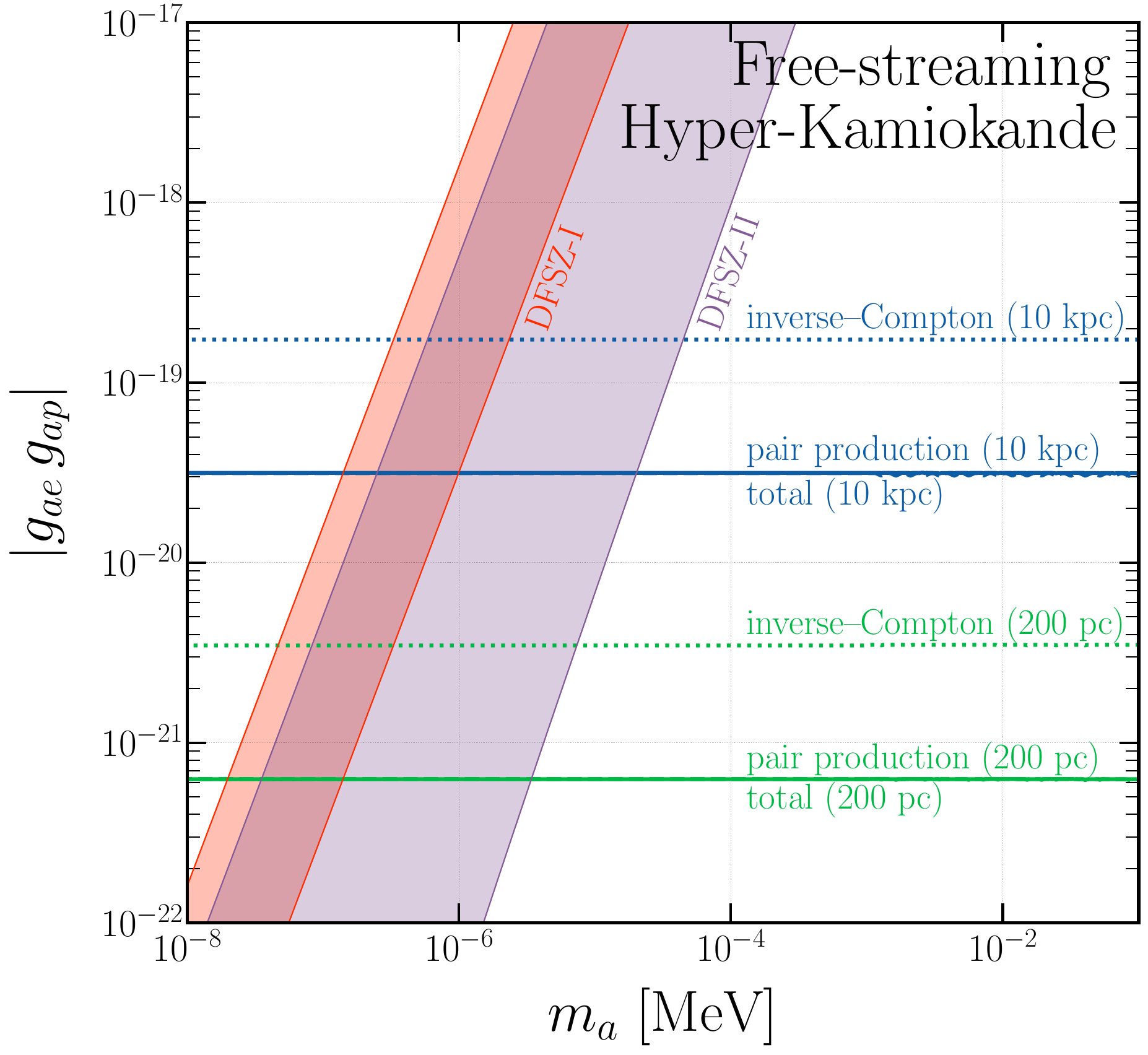}
\caption{Exclusion region plot in the $(|g_{ap}\, g_{ae}|,\,m_a)$ plane at $90\%$ C.L. for the free-streaming regime. The left panel refers to the sensitivity of JUNO, while the right one to the sensitivity of Hyper-K. The blue lines represent the sensitivity for axions produced in a SN at $10$~kpc, while the green one from a SN at $200$~pc. The dotted lines denotes the sensitivity from inverse Compton scattering, the dashed ones from axion-induced external pair production and the solid ones from their combination. The red and purple bands show the preferred regions for the DFSZ-I and DFSZ-II models.}
\label{fig:SN_FS_final}       % Give a unique label
\end{figure*}

\begin{figure*}[t!]
\centering
\includegraphics[width=0.49 \textwidth,clip]{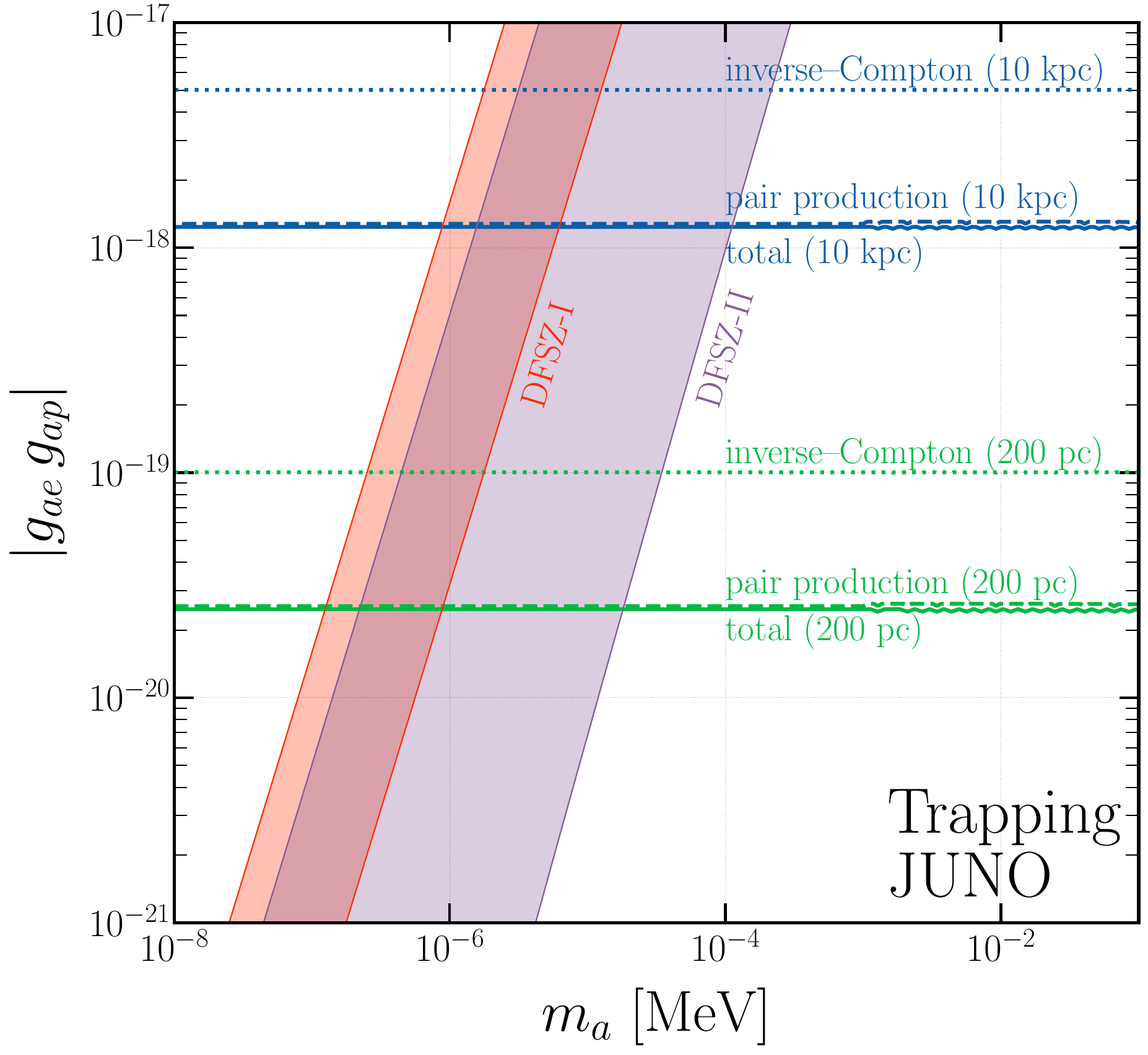}
\includegraphics[width=0.49 \textwidth,clip]{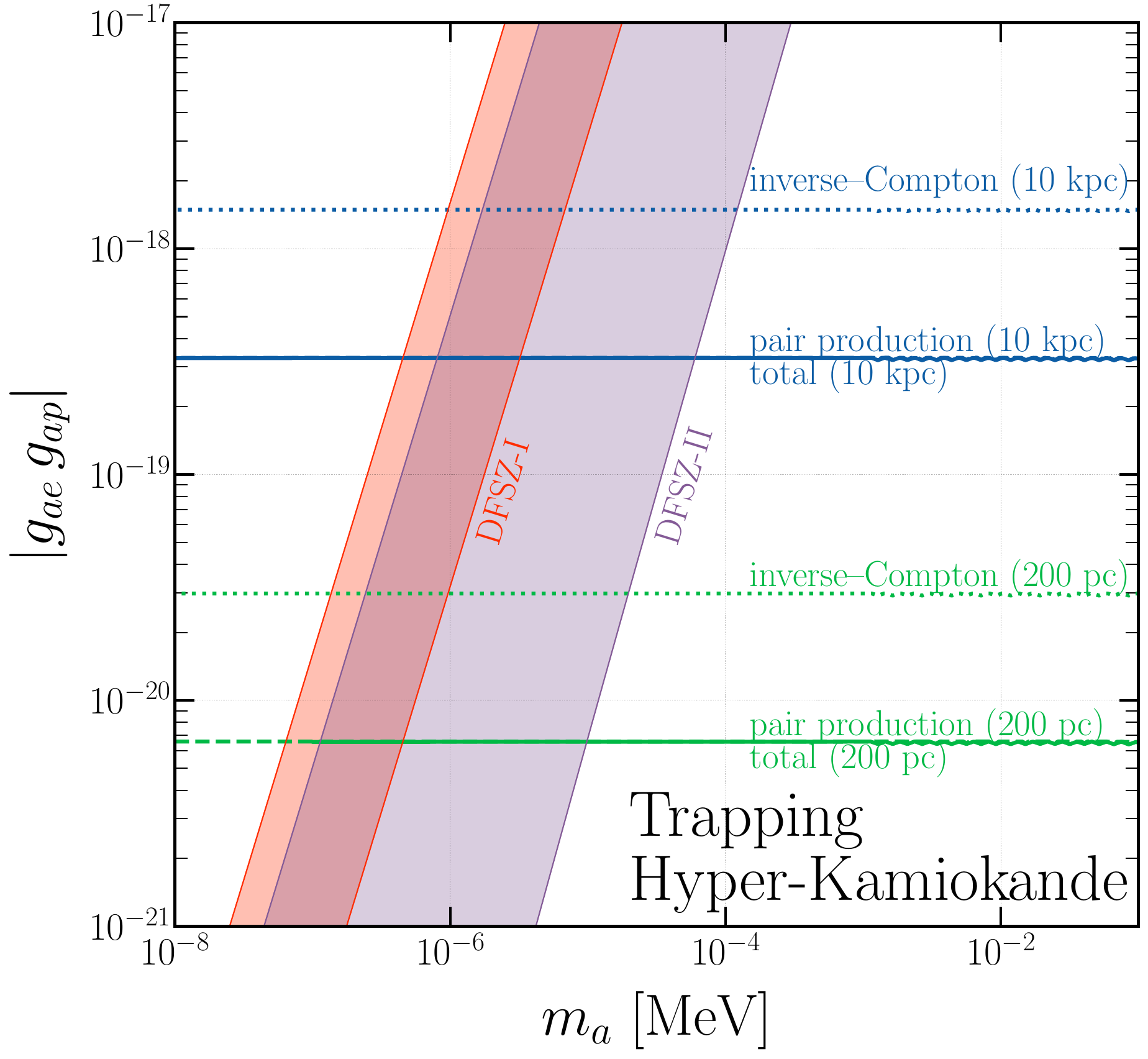}
\caption{Exclusion region in the $(|g_{ap}\,g_{ae}|,\,m_a)$ plane at $90\%$ C.L. for the trapping regime. The left panel refers to the sensitivity of JUNO, while the right one to the sensitivity of Hyper-K. The blue lines represent the sensitivity for axions produced in a SN at $10$~kpc, while the green one from a SN at $200$~pc. The dotted lines denotes the sensitivity from inverse Compton scattering, the dashed ones from axion-induced external pair production and the solid ones from their combination. The red and purple bands show the preferred regions for the DFSZ-I and DFSZ-II models.}
\label{fig:SN_Tr_final}       % Give a unique label
\end{figure*}

In this section, we estimate the sensitivity of underground detectors to SN axions, including external pair production.\footnote{To estimate the sensitivity, we used the same efficiency and fiducial volume as those employed in solar axion searches.} 
A summary of our findings is presented in Fig.~\ref{fig:SN_FS_final}, Fig.~\ref{fig:SN_Tr_final}, and in Tab.~\ref{tab:Summary}.

\vspace{0.2cm}
\noindent
\textbf{Free-streaming regime.} 
At JUNO, the expected  number of events from inverse Compton scattering with an energy larger than $E_\mathrm{cut-off}=100$ MeV due to axions produced via nucleon-nucleon Bremsstrahlung and pionic Compton scattering in the free-streaming regime (in the $10$ s of the burst) from a SN at $10$ kpc is\footnote{Axions from SN impose limits on the $|g_{ae}\,g_{ap}|$ couplings, where, from Eq.~\eqref{eq:Lagrangian_int}, $g_{ap} = g_{0\mathrm{aN}}-g_{3\mathrm{aN}}$.}
\begin{equation}
    S_\mathrm{expected} \simeq 8.7\times 10^{36} g_{ap}^2\,g_{ae}^2,
\end{equation} 
providing a sensitivity of $|g_{ap}\,g_{ae}| \lesssim 5.9 \times 10^{-19}$. For the external pair production in the field of nuclei, the expected event count is
\begin{equation}
    S_\mathrm{expected} \simeq 2.0 \times 10^{38} g_{ap}^2g_{ae}^2,
\end{equation}
yielding a sensitivity of $|g_{ap}\,g_{ae}| \lesssim 1.2\times 10^{-19}$. Combining these processes slightly enhances the expected number of events 
\begin{equation}
    S_\mathrm{expected} \simeq 2.1\times 10^{38} g_{ap}^2g_{ae}^2,
\end{equation}
resulting in $|g_{ap}\,g_{ae}| \lesssim 1.2\times 10^{-19}$.

The expected number of events scales as the square of the distance from the SN. For example, for a relatively close SN, such as Betelgeuse at $\sim200$ pc, the expected events from inverse Compton scattering are
\begin{equation}
    S_\mathrm{expected} \simeq 2.1\times 10^{40} g_{ap}^2g_{ae}^2,
\end{equation}
providing a sensitivity down to $|g_{ap}\,g_{ae}| \sim 1.2\times 10^{-20}$. 
For pair production alone, we find 
\begin{equation}
    S_\mathrm{expected} \simeq 5.1\times 10^{41} g_{ap}^2g_{ae}^2,
\end{equation}
resulting in sensitivity to $|g_{ap}\,g_{ae}| \sim 2.4\times 10^{-21}$.
Combining both processes yields
\begin{equation}
    S_\mathrm{expected} \simeq 5.3\times 10^{41} g_{ap}^2g_{ae}^2,
\end{equation}
giving sensitivity down to $|g_{ap}\,g_{ae}| \sim 2.4\times 10^{-21}$.

These results scale with distance, allowing sensitivity estimates for any SN distance via simple rescaling:
\begin{equation}
    |g_{ap}\,g_{ae}|(d) = |g_{ap}\,g_{ae}|(10\,\mathrm{kpc})\, \frac{d}{10\,\mathrm{kpc}}.
\end{equation}

The left side of Fig.~\ref{fig:SN_FS_final} shows results for SN axions in the free-streaming regime for JUNO, with benchmarks at $10$ kpc and Betelgeuse's $\sim200$ pc. The plot displays $|g_{ae}\,g_{ap}|$ as a function of $m_a$. Blue curves represent the potential reach for a 10 kpc SN, with dotted, dashed and solid lines indicating the sensitivity for inverse Comptons scattering, pair production and their combination, respectively.
Green curves shows results for Betelgeuse. 
As expected from Fig.~\ref{fig:xscomparison}, external pair production is the dominant detection channel for energetic SN axions in the free-streaming regime, given the cut-off energy of $100$ MeV to suppress the SN neutrino background.

At Hyper-K, the bounds improve by about a factor of $4$ with respect to the one from JUNO and are presented in Tab.~\ref{tab:Summary} and in the right panel of Fig.~\ref{fig:SN_FS_final}. Hyper-K gives a stronger bound because of its largest fiducial volume compared to the one of JUNO, while the worst energy resolution does not play a role in SN axion searches.

\vspace{0.2cm}
\noindent
\textbf{Trapping regime.} In the trapping regime, for JUNO the expected number of events from inverse Compton scattering alone above $E_\mathrm{cut-off}$ for a SN at $10$ kpc is
\begin{equation}
    S_\mathrm{expected} \simeq 1.2\times10^{35} g_{ap}^2\,g_{ae}^2,
\end{equation}
leading to a sensitivity of $|g_{ap}\,g_{ae}| \sim 5.0\times 10^{-18}$. 
For pair production, we find
\begin{equation}
    S_\mathrm{expected} \simeq 1.8\times10^{36} g_{ap}^2\,g_{ae}^2,
\end{equation}
with sensitivity down to $|g_{ap}\,g_{ae}| \sim 1.5\times 10^{-18}$. 
Combining both detection channels yields
\begin{equation}
    S_\mathrm{expected} \simeq 2.0 \times10^{36} g_{ap}^2\,g_{ae}^2,
\end{equation}
with a sensitivity of $|g_{ap}\,g_{ae}| \sim 1.2\times 10^{-18}$.

For Betelgeuse, the expected events for inverse Compton scattering are 
\begin{equation}
    S_\mathrm{expected} \simeq 3.0 \times10^{38} g_{ap}^2\,g_{ae}^2
\end{equation}
yielding $|g_{ap}\,g_{ae}| \sim 1.0\times 10^{-19}$,
while combining it with pair production  leads to \begin{equation}
    S_\mathrm{expected} \simeq 4.9 \times10^{39} g_{ap}^2\,g_{ae}^2
\end{equation}
with sensitivity reaching $|g_{ap}\,g_{ae}| \sim 2.5 \times 10^{-20}$.

Figure~\ref{fig:SN_Tr_final} shows the sensitivity in the $(|g_{ap} g_{ae}|,\,m_a)$ plane at $90\%$ C.L. in the trapping regime. The  blue curves indicate sensitivity for a SN at $10$ kpc, while the green curves corresponds to a closer SN at $200$ pc. The left panel shows the results for JUNO.
Hyper-Kamiokande will reach a factor of $\sim4$ better sensitivity with respect to JUNO also in the trapping regime. The corresponding results are shown in Tab.~\ref{tab:Summary} and Fig.~\ref{fig:SN_Tr_final}.

\begin{table*}[]
\centering
\label{tab:Summary}
\renewcommand{\arraystretch}{1.8}
\begin{tabular}{c|cc|cc|cc|}

\cline{2-7} & \multicolumn{2}{P{4cm}|}{$|g_{3aN}\,g_{ae}|$ -- 5.5 MeV Line}    & \multicolumn{2}{P{4.6cm}|}{$|g_{ap}\,g_{ae}|$ -- SN (Free Streaming)} & \multicolumn{2}{P{4.2cm}|}{$|g_{ap}\,g_{ae}|$ -- SN (Trapping)}  \\
\hline \multicolumn{1}{|c|}{Pair Production} & \multicolumn{1}{P{2cm}|}{No} & \multicolumn{1}{P{2cm}|}{Yes}     & \multicolumn{1}{P{2.3cm}|}{No} & \multicolumn{1}{P{2.3cm}|}{Yes} & \multicolumn{1}{P{2.1cm}|}{No} & \multicolumn{1}{P{2.1cm}|}{Yes} \\ \hline
\multicolumn{1}{|c|}{Borexino} & \multicolumn{1}{c|}{$\lesssim5.7\times 10^{-13}$} & $\lesssim4.8\times 10^{-13}$ &\multicolumn{1}{c|}{--} &\multicolumn{1}{c|}{--} &\multicolumn{1}{c|}{--} &\multicolumn{1}{c|}{--} \\ \hline
\multicolumn{1}{|c|}{JUNO} & \multicolumn{1}{c|}{$\lesssim6.6\times 10^{-14}$}      & $ \lesssim 5.6 \times 10^{-14}$ & \multicolumn{1}{c|}{$\lesssim 5.9\times 10^{-19}$}   & \multicolumn{1}{c|}{$\lesssim 1.2\times 10^{-19}$}& \multicolumn{1}{c|}{$\lesssim 5.0\times 10^{-18}$}& \multicolumn{1}{c|}{$\lesssim 1.2\times 10^{-18}$} \\ \hline
\multicolumn{1}{|c|}{Hyper-K} & \multicolumn{1}{c|}{$\lesssim1.1\times 10^{-13}$} & $ \sim 9.1 \times 10^{-14}$ & \multicolumn{1}{c|}{$\lesssim 1.7\times 10^{-19}$} & \multicolumn{1}{c|}{$\lesssim3.1\times10^{-20}$}& \multicolumn{1}{c|}{$\lesssim 1.5\times10^{-18}$}& \multicolumn{1}{c|}{$\lesssim 3.3\times 10^{-19}$} \\ \hline
\end{tabular}
\caption{Summary results for the coupling $|g_{3aN}\,g_{ae}|$ and $|g_{ap}\,g_{ae}|$ respectively from solar and SN axions (assuming a 10 kpc distance), with and without the inclusion of the axion pair production process. The entries indicate either bounds, when explicit measurements have been taken, or sensitivity prospects.\label{tab:Summary}}
\end{table*}

\section{Conclusions}
\label{sec:conclusions}
In this work, we explored axion detection through the process of axion-induced pair production in a nuclear electric field. This process represents the axion equivalent of the well-known Bethe-Heitler process~\cite{Bethe:1934za}, whose application to axion searches has been largely overlooked in previous literature. We provide a new derivation of this process, including the nuclear form factor, and present a simple expression for the corresponding cross section in terms of the Bethe-Heitler cross section, which can be readily applied to specific experimental scenarios.

Our study demonstrates that this mechanism offers competitive sensitivity compared to traditional detection channels, especially for highly energetic ALPs. In particular, our analysis shows that axion pair production dominates over the inverse Compton scattering process at energies above approximately 10~MeV, making it an optimal approach for detecting axions or ALPs produced in nuclear reactions or supernova (SN) explosions (see Fig.~\ref{fig:xscomparison}). We have applied this mechanism to two scenarios: the 5.5 MeV solar axion line produced in the solar pp-chain, and SN axions, assuming a galactic SN event.

In exploring these cases, we analyzed the sensitivity of established underground experiments, such as Borexino, JUNO, and Hyper-K, to these axions, comparing scenarios with and without the inclusion of the pair production process. Our results, summarized in Tab.~\ref{tab:Summary}, show how incorporating axion pair production significantly enhances sensitivity and improves the bounds, particularly for SN axions. This is evident also from Figs.~\ref{fig:solar_final}, \ref{fig:SN_FS_final}, and \ref{fig:SN_Tr_final}, which present our findings for the solar and SN cases respectively.

Notice that an experiment like Borexino would not be competitive on SN axion searches with respect to JUNO and Hyper-K because of the smaller fiducial volume. However, the better energy resolution makes its sensitivity comparable to that of the next generation experiments for solar axion searches.

By demonstrating the potential impact of this mechanism on detection capabilities, particularly for SN axions with energies above a few MeV, we underline the need to consider axion pair production as a viable detection channel in future experimental searches.

These findings highlight the importance of including axion pair production in the broader context of axion and ALP detection, as it may not only enable new constraints on axion properties but also offer deeper insights into stellar environments and fundamental physics.

\section*{Acknowledgments}
We warmly thank Thomas Janka for providing access to the {\tt GARCHING} group archive. We also acknowledge Andrea Longhin, Alessandro Paoloni, and Benjamin Quilai for clarifying the sensitivities of the JUNO and Hyper-K experiments. 
Additionally, we appreciate valuable 
discussions with Ralph Massarczyk, Christoph Ternes and Francesco Vissani. 
The work of FAA and FM is supported
by the European Union -- Next Generation EU and
by the Italian Ministry of University and Research (MUR) 
via the PRIN 2022 project n.~2022K4B58X -- AxionOrigins. FAA, FM and GGdC are also supported by the INFN “Iniziativa Specifica” Theoretical Astroparticle Physics (TAsP). MG acknowledges support from the Spanish Agencia Estatal de Investigación under grant PID2019-108122GB-C31, funded by MCIN/AEI/10.13039/501100011033, and from the “European Union NextGenerationEU/PRTR” (Planes complementarios, Programa de Astrofísica y Física de Altas Energías). He also acknowledges support from grant PGC2022-126078NB-C21, “Aún más allá de los modelos estándar,” funded by MCIN/AEI/10.13039/501100011033 and “ERDF A way of making Europe.” Additionally, MG acknowledges funding from the European Union’s Horizon 2020 research and innovation programme under the European Research Council (ERC) grant agreement ERC-2017-AdG788781 (IAXO+). 
This article/publication is based upon work from COST Action COSMIC WISPers CA21106, supported by COST (European Cooperation in Science and Technology).

\onecolumngrid
\appendix 

\section{The axion--induced external pair production cross section}
\label{app:full}

The process 
$a(k_a) + {}^A_ZX(p_1) \to {}^A_ZX(p_2) + e^+(\ell_+) + e^-(\ell_-)$ depicted in Fig.~\ref{fig:FeynPairProduction} is the analogous of the photon-lepton pair production in the Standard Model, the so called Bethe-Heitler process. It was computed for the first time in Ref.~\cite{Kim:1984ss, Kim:1982xb}, using the atomic form factor of Ref.~\cite{Tsai:1973py}.
We use the following definitions for the four-momenta of the process: $k_a=(E_a,\vec{k}_a)$ is the axion 4-momentum, $p_1=(m_X,0)$ is the initial state target 4-momentum, $p_2=(E_2, \vec{p_2})$ is the final state target  4-momentum, $\ell_+=(E_+,\vec{\ell_+})$ is the positron 4-momentum and $\ell_-=(E_-,\vec{\ell_-})$ is the electron 4-momentum.
Following the above definitions we have:
\ba
    k_a\cdot \ell_\pm &= E_a E_\pm - \vec{k}_a\cdot \vec{\ell}_\pm \\
    k_a\cdot p_1 &= E_a m_X\\
    k_a\cdot p_2 &= E_a E_2  - \vec{k}_a\cdot \vec{p}_2 = E_a \sqrt{|\vec{p}_2|^2 + m_X^2} - \vec{k}_a\cdot\vec{p}_2 \simeq E_a m_X - \vec{k}_a\cdot\vec{p}_2\\
    \ell_+ \cdot \ell_- &= E_+ E_- - \vec{\ell}_+ \cdot \vec{\ell}_-\\
    \ell_{\pm} \cdot p_1 &= E_\pm m_X\\
    \ell_\pm \cdot p_2 &= E_\pm E_2 - \vec{\ell}_\pm \cdot \vec{p}_2
\ea
We take a reference frame where $\vec{k}_a=(0,0,|\vec k_a|)$,  $\vec{\ell}_+=|\ell_+|(\sin\theta_+ \cos\phi_+, \sin\theta_+ \sin\phi_+,\cos\theta_+)$ and $\vec{\ell}_-=|\ell_-|(\sin\theta_- \cos\phi_-, \sin\theta_- \sin\phi_-,\cos\theta_-)$. Furthermore, we define the angle $\phi=\phi_+-\phi_-$ such that $\phi$ is the angle between the planes $\vec{k}_a\cdot \vec{\ell}_+$ and $\vec{k}_a\cdot \vec{\ell}_-$. As a consequence, we have 
\ba
    \vec{\ell}_\pm\cdot\vec{k}_a &= |\vec k_a||\vec{\ell}_\pm|\cos\theta_\pm\\
    \vec{\ell}_+ \cdot \vec{\ell}_- &= |\vec{\ell}_+||\vec{\ell}_-| \biggl[ \sin\theta_+ \sin\theta_- (\cos\phi_+\cos\phi_-  + \sin\phi_+ \sin\phi_-) + \cos\theta_+ \cos\theta_- \biggr]\\
    &= |\vec{\ell}_+||\vec{\ell}_-| \biggl[ \sin\theta_+ \sin\theta_- \cos\phi +
    \cos\theta_+ \cos\theta_- \biggr]\\
\vec{k}_a\cdot \vec{p_2} &= |\vec k_a|^2 - |\vec k_a||\vec{\ell}_+| \cos\theta_+ -|\vec k_a||\vec{\ell}_-| \cos\theta_- \\
    \vec{\ell}_\pm\cdot \vec{p_2} &=  |\vec k_a||\vec{\ell}_\pm|\cos\theta_\pm -
    |\vec{\ell}_+||\vec{\ell}_-| \frac{1}{4}\biggl( \sin2\theta_+ \sin2\theta_- \cos\phi  \biggr) - |\vec{\ell}_\pm|^2
\ea
Now we derive the differential cross section:
\ba
    d\sigma &= \frac{(2\pi)^4|\mathcal{M}|^2 \delta^{(4)}(k_a+p_1-q)}{4\sqrt{(k_a\cdot p_1)^2 - m_a^2 m_X^2}}
    d\Phi^3
    = \frac{|\mathcal{M}|^2\delta(E_a+E_1-E_q)}{4\sqrt{(k_a\cdot p_1)^2 - m_a^2 m_X^2}} \frac{d^3\ell_+ d^3\ell_-}{8 (2\pi)^5 E_+E_-E_2}\\
    &= \frac{|\mathcal{M}|^2\delta(E_a+E_1-E_q)}{32 (2\pi)^5 m_X |\vec k_a|}\frac{d^3\ell_+d^3\ell_-}{E_+E_-E_2}
\ea
where $d\Phi^3=\prod_{F=+,-,2}d^3 p_F/(2\pi)^3 2 E_F$ is the three-body phase
with $q=\ell_++\ell_-+p_2$ and $E_q=E_{+}+E_{-}+E_2$.
Now we expand the $d^3\ell_i$ and take into account that the momentum of the target in the final state is much smaller than the mass of the target in the final state: $E_2\sim m_X=E_1$. We have then:
\ba
    d\sigma &= \frac{|\mathcal{M}|^2\delta(E_a -E_+-E_-)}{32(2\pi)^5 m_X^2 |\vec k_a| E_+E_-} d\cos\theta_+ |\vec{\ell}_+|^2 d\ell_+ d\phi_+ d\cos\theta_- |\vec{\ell}_-|^2 d\ell_- d\phi_- \\
    &= \frac{|\mathcal{M}|^2\delta(E_a -E_+-E_-)}{32(2\pi)^5 m_X^2 |\vec k_a|} d\cos\theta_+ |\vec{\ell}_+| d E_+ d\phi_+ d\cos\theta_- |\vec{\ell}_-| d E_- d\phi_-\\
    &= \frac{|\mathcal{M}|^2 d\cos\theta_+ |\vec{\ell}_+| d\phi_+ d\cos\theta_- |\vec{\ell}_-| d E_- d\phi_- }{32(2\pi)^5 m_X^2 |\vec k_a| }\\
    &= \mathcal{J} \frac{|\mathcal{M}|^2   |\vec{\ell}_+|   |\vec{\ell}_-| d\cos\theta_+ d\cos\theta_- d E_- d\phi }{32(2\pi)^4 m_X^2 |\vec k_a| } \\
    &= \mathcal{J}  \frac{|\mathcal{M}|^2  \, |\vec{\ell}_+| \, |\vec{\ell}_-|\, d\cos\theta_+\, d\cos\theta_- \,d E_- \, d\phi }{512\, \pi^4\, m_X^2\, |\vec k_a| }
\ea
where we used the fact that $\ell d\ell = E dE$, integrated in $dE_+$ using the delta function that imposes $E_+ = E_- - E_a$, and $\mathcal{J} = -1$ is the Jacobian for the transformation of the $\phi$ angle.
The matrix element modulo squared in Eq.~(\ref{eq:Mq}) is in details 

\ba
    |\mathcal{M}|^2 &= \frac{e^4 g_{ae^2}}{t^2}(|\mathcal{M}_1|^2 + |\mathcal{M}_2|^2 + 2\,\mathcal{M}_{12})\\
    |\mathcal{M}_1|^2 &= \frac{32}{(m_a^2 - 2(k_a\cdot \ell_+))^2} \biggl[ 2 m_X^2 (k_a\cdot \ell_-)(k_a\cdot \ell_+) - 2 (k_a\cdot \ell_+) (k_a\cdot p_2) (p_1\cdot \ell_-) \\
    &- 2 (k_a\cdot \ell_+) (k_a\cdot p_1) (p_2 \cdot \ell_-) + m_a^2 \biggl( - m_X^2 (\ell_+ \cdot \ell_-) + (p_2\cdot \ell_-) (p_1\cdot \ell_+) \\
    &+ (p_2\cdot \ell_+) (p_1\cdot \ell_-) + m_e^2 ((p_1\cdot p_2) - 2 m_X^2) \biggr)\biggr]\\
    |\mathcal{M}_2|^2 &= \frac{32}{(m_a^2-2 (  k_a\cdot \ell_-))^2} \biggl[2 (  k_a\cdot \ell_-) \biggl(m_X^2 (  k_a\cdot \ell_+)-(  k_a\cdot p_2) (  \ell_+\cdot p_1)  \\
    &-(  k_a\cdot p_1) (  \ell_+\cdot p_2)\biggr) + m_a^2 \biggl(- m_X^2 (\ell_- \cdot \ell_+)+(  \ell_-\cdot p_2) (  \ell_+\cdot p_1) \\
    &+ (  \ell_-\cdot p_1) (  \ell_+\cdot p_2)+m_e^2 ((  p_1\cdot p_2)-2 m_X^2)\biggr)\biggr]\\
    \mathcal{M}_{12} &= -\frac{32}{(m_a^2-2 (k_a \cdot \ell_-)) (m_a^2-2 (k_ \cdot \ell_+))} \biggl((k_a \cdot \ell_-) \biggl(-2 m_X^2 (k_a \cdot \ell_+)\\
    &+ (k_a \cdot p_2) (\ell_+ \cdot p_1)+(k_a \cdot p_1) (\ell_+ \cdot p_2)\biggr)-2 (k_a \cdot p_1) (k_a \cdot p_2) \biggl((\ell_- \cdot \ell_+)+m_e^2\biggr)\\
    &+ (k_a \cdot \ell_+) \biggl((k_a \cdot p_2) (\ell_- \cdot p_1)+(k_a \cdot p_1) (\ell_- \cdot p_2)\biggr)-m_a^2 \biggl((\ell_- \cdot p_2) (\ell_+ \cdot p_1) \\
    &+(\ell_- \cdot p_1) (\ell_+ \cdot p_2)-(\ell_- \cdot \ell_+) (p_1 \cdot p_2)\biggr)+m_X^2 m_a^2 m_e^2\biggr)
\ea

\bibliographystyle{apsrev4-1.bst}
\bibliography{references.bib}

\end{document}